\documentclass[acmsmall, authorversion]{acmart}
\usepackage[utf8]{inputenc}

\setcopyright{acmlicensed}
\acmJournal{PACMHCI}
\acmYear{2021} \acmVolume{5} \acmNumber{CSCW2} \acmArticle{371} \acmMonth{10} \acmPrice{15.00}\acmDOI{10.1145/3479515}

\begin{document}

\author{William Seymour}
\email{william.seymour@cs.ox.ac.uk}
\orcid{0002-0256-6740}
\affiliation{
  \institution{University of Oxford}
  \streetaddress{Parks Road}
  \city{Oxford}
  \country{UK}
  \postcode{OX1 3QD}
}
\authornote{Now at King's College London: william.1.seymour@kcl.ac.uk, Bush House, 30 Aldwych, London, WC2B 4BG}

\author{Max Van Kleek}
\email{max.van.kleek@cs.ox.ac.uk}
\affiliation{
  \institution{University of Oxford}
  \streetaddress{Parks Road}
  \city{Oxford}
  \country{UK}
  \postcode{OX1 3QD}
}



\title[Exploring Interactions Between Trust, Anthropomorphism, and Relationship Development in Voice Assistants]{Exploring Interactions Between Trust, Anthropomorphism, and Relationship Development in Voice Assistants}

\begin{abstract}
Modern conversational agents such as Alexa and Google Assistant represent significant progress in speech recognition, natural language processing, and speech synthesis. But as these agents have grown more realistic, concerns have been raised over how their social nature might unconsciously shape our interactions with them. Through a survey of 500 voice assistant users, we explore whether users' relationships with their voice assistants can be quantified using the same metrics as social, interpersonal relationships; as well as if this correlates with how much they trust their devices and the extent to which they anthropomorphise them. Using Knapp's staircase model of human relationships, we find that not only can human-device interactions be modelled in this way, but also that relationship development with voice assistants correlates with increased trust and anthropomorphism.
\end{abstract}

\maketitle

\section{Introduction}
\label{sec:intro}
In the 1927 silent film \textit{Metropolis}, the inventor Rotwang creates a `maschinenmensch' to resurrect his dead lover~\cite{lang1927metropolis}. One of the first robots to ever appear in film, its human appearance is used to manipulate relationships amongst the main characters and start an uprising. Several decades later, Asimov coined the term robotics and predicted the social and emotional challenges of interacting with realistic `humaniform' robots~\cite{asimov1994robots}. More recent films such as \textit{Her} and \textit{Ex Machina} have continued to explore relationships with machines that look and/or sound human in a modern context~\cite{garland2014ex, jonze2013her}.

In the 1990's, Nass et al. showed that we are unable to engage rationally with voice interfaces; speech is so tightly integrated with many interpersonal functions in the brain that we cannot help but respond socially to devices with voices~\cite{nass1994computers, nass2005wired, reeves1996media}. Weizenbaum had already said of ELIZA ``how easy it is to create and maintain the illusion of understanding [...] A certain danger lurks there''~\cite{weizenbaum1966eliza}, and Rosalind Picard was cognisant of the capability of affective computers to mislead~\cite{picard2000affective}. Despite this, it was taken for granted in the early days of conversational agents that human-like, embodied agents were the ``logical extension of the metaphor of human--computer interaction as a conversation''~\cite{cassell1999embodiment}. It seemed obvious to make use of communication skills that people already possessed, as well as the fact that natural language has built in redundancy and can be supplemented by gestures, tone, and facial expressions. On the face of it, these interfaces represented an ideal `invisible' human-computer interface.

The disembodied voice assistants commercially available today---such as Amazon Alexa and Google Assistant---are sophisticated conversational agents, leveraging advances in speech recognition, natural language processing, and speech synthesis. The widespread adoption of these products has foregrounded this natural and unconscious treatment of conversational agents as social actors, and has brought with it a host of concerns as people integrate these social devices into their daily lives.

This would suggest that the relationships that we have with voice assistants are inherently different from those that we have with other smart technologies~\cite{turkle2007authenticity}, falling somewhere between the clear-cut categories of human and machine. But while there have been many contributions investigating how people use voice assistants, little work has been done to explore the \textit{nature} of our relationships with them, including the potential cognitive shifts that come about when conversing with agents that present themselves as human. Given that these agents rely on the same social skills utilised in developing interpersonal relationships\footnote{Relationship here refers to the level and nature of communication between two parties. Relationships that are closer or more developed tend to feature higher levels of intimacy and self disclosure~\cite{altman1973social}.}, it is important that we better understand the similarities and differences between human relationships with voice assistants and relationships with other humans. And if this comparison \textit{does} makes sense, then how do human-assistant relationships relate to other interaction components that are already well understood, such as trust and anthropomorphism? To this end, we pose and answer the following research questions about how people interact with voice assistants:

\begin{enumerate}
    \item [$RQ_1$] Do users of voice assistants exhibit relationship development with their devices similar to that measured between humans?
    \item [$RQ_2$] Are there correlations between the extent to which users trust, develop relationships with, and anthropomorphise voice assistants?
    \item [$RQ_3$] Are there secondary characteristics (ownership duration, being the primary account holder, etc.) that correlate with trust, anthropomorphism, and/or relationship development in voice assistants?
\end{enumerate}

In doing so we make the following contributions:
\begin{itemize}
    \item We explore the potential of relationship development between users and voice assistants as a new way of conceptualising people's interactions with such devices, identifying correlations between how much participants trusted, anthropomorphised, and developed relationships with their voice assistants
    \item We show that some secondary characteristics of voice assistant ownership are similarly correlated with higher trust and/or relationship development
    \item We discuss these correlations in the context of prior work on smart devices and interpersonal interaction, including how voice assistants trigger shifts in explanatory categories, rational distributions of trust, and the potential for voice assistants to be emotionally manipulative
\end{itemize}

\section{Background}
Human relationships have been the focus of intense study for centuries, with psychologists analysing the social mechanisms that pull us together and encourage mutual trust, information sharing, and intimacy. More recently, HCI research has shown that computers are social actors in ways that are similar to, but distinct from, humans. Building on this, the Human-Robot Interaction (HRI) community has explored how people interact with social robots, utilising custom hardware to explore the effects of varying physical forms on participant response~\cite{bartneck2009measurement, fussell2008people, hancock2011meta, salem2015would}. A burgeoning body of work on voice assistants has also begun to study device usage and integration \textit{in situ}~\cite{abdi2019more, bentley2018understanding, lopatovska2018personification, porcheron2018voice}. In this section we outline ongoing research efforts across these diverse fields in order to situate the results presented in this study.

\subsection{Human Interpersonal Relationships}
A large body of psychology and communications theory literature is dedicated to understanding and classifying facets of human interaction. While many theories of interpersonal interaction focus on experiential aspects of interpersonal relationships that are potentially less relevant to smart devices (such as sexual desire or interdependence~\cite{chelune1984cognitive, davis1982friendship}), others have proposed `stage' models of relationships that encompass the behavioural aspects of relationships from meeting to parting. Murstein~\cite{murstein1970stimulus}, Altman and Taylor~\cite{altman1973social}, and Knapp and Vangelisti~\cite{knapp1978social, knapp2005interpersonal}, each proposed stage theories of relationships, suggesting that relationships develop up (and down) identifiable strata of feelings and behaviour. These behavioural signals help to capture the essence of these relationships in ways that ambiguous self-defined labels such as `friend' or `partner' can struggle to~\cite{wright1978toward}.

While Murstein's work pertains more to partner selection, Altman and Taylor's work on the process of social penetration conceptualises wider relationships in terms of breadth and depth of interaction~\cite{altman1973social}. As individuals progress along a relationship trajectory of four stages they move along both of these axes to varying extents. Attempts to operationalise these breadth and depth concepts typically involve rating different disclosure topics for intimacy~\cite{taylor1966intimacy}. Knapp's staircase model of relationships also documents the shift in interpersonal communication as a relationship develops, and integrates a number of core ideas from social penetration theory~\cite{knapp1978social, knapp2005interpersonal}. The staircase model proposes five stages of both development (from \textit{initiating} to \textit{bonding}) and parting (from \textit{differentiating} to \textit{terminating}) that describe communication patterns within relationships. As with social penetration theory, movement through stages can be bidirectional and is linear within meeting/parting trajectories. After initial work by Avtgis et al.~\cite{avtgis1998relationship}, Welch and Rubin later adapted Knapp's model to identify \textit{behaviours}, held to be more replicable than emotions (e.g. using the extent to which individuals share information as a partial proxy for intimacy)~\cite{welch2002development}. In Section~\ref{sec:method} we adapt a selection of these questions to the context of voice assistants through a series of pilot studies.

\subsection{Computers as Social Actors}\label{personification}
Pioneering work by Nass et al. in 1994, of what would later become  the Computers are Social Actors (CASA) paradigm, showed that people sometimes applied social rules (and took corresponding social actions) when interacting with computers even whilst being fully aware that such computers were machines, not humans~\cite{nass1994computers}. This has been described as a kind of cognitive dissonance (see~\cite{fiske2013social}), in which individuals experienced difficulty rationalising their actions with computers during Nass' experiments. Further supporting this view were a series of experiments carried out by Nass, Steuer, and Tauber, which demonstrated that people: (1) were prone to apply notions of `self' and `other' in interactions with computers; (2) drew upon and applied gender stereotypes in such interactions; and, finally, (3) exhibited automatic and unconscious social responses in particular settings~\cite{nass1994computers} (see also~\cite{nass2005wired, reeves1996media}, later replicated with voice assistants in~\cite{lopatovska2018personification}).

This same methodology was used by Moon and Nass to show that social routines such as reciprocal information sharing~\cite{moon2000intimate} and mindless re-enactment of existing social scripts~\cite{nass2000machines} also apply to human-computer interactions. This mindless interaction is reminiscent of overlearned social routines where any task-components that seem applicable are blindly accepted as true~\cite{fiske2013social}. Early work on affective computing also recognised the potential for machines to manipulate and deceive using emotions, although it did not anticipate this affecting subsequent interpersonal interactions~\cite{picard2000affective, picard2003affective}.

A major discussion that arises around conversational voice interfaces is therefore the extent to which they should emulate the sounds, structure, and mannerisms of human speech. Human voices largely outperform synthesised ones in preference and quality tests~\cite{cambre2020choice}, but there has been significant criticism of the performance of gender by voice assistants~\cite{aylett2019siri, loideain2020alexa, walker2020alexa, west2019blush}\footnote{The Q genderless voice represents an interesting development in this area: \url{https://www.genderlessvoice.com}}, and the recent release of Google Duplex, an AI assistant that can make phone calls using convincing human speech\footnote{\url{https://ai.googleblog.com/2018/05/duplex-ai-system-for-natural-conversation.html}}, has reinvigorated discussions about the extent to which voice interfaces should be able to masquerade as human. 

Work on the design of voice assistants has recognised the ``need to reconsider symbiosis between humans and AI---as partners rather than a user and a tool''~\cite{cho2020role}, with personification predicting user satisfaction even in failed interactions~\cite{lopatovska2018personification, purington2017alexa}. Other work has explored the effects of voice assistants on the social order of the home~\cite{porcheron2018voice}, how older adults relate to voice assistants~\cite{pradhan2019phantom}, and the effects of modality on information disclosure~\cite{cho2019hey}. There has been some trouble, however, in determining the extent to which these results represent a shift towards deeper social relationships vs mindless behaviours triggered by the use of speech as a modality; observed interactions with voice assistants seem to sit between traditional human-computer and human-human interactions. This work therefore seeks to address this gap in existing literature, extending our knowledge on the nature of our relationship with voice enabled devices and developing the discussion around how human-like they should be designed to be.

\subsection{Embodiment and Anthropomorphism}
People have long sought to embody computers in lifelike guises. The most visible examples are in works of science fiction, but HRI has a long history of exploring the possibilities of embodied robots; Cynthia Breazeal's Kismet~\cite{breazeal2004designing} and the Paro therapeutic robot\footnote{\url{http://www.parorobots.com/}}, for example, demonstrate the social potential of robots that take on animal forms. While the nature of an agent's embodiment obviously has an impact on what it can sense about the world, the extra information present in an embodied robot is also used by people to generate mental models about how the robot functions~\cite{lee2005human, zawieska2012understanding}.

Anthropomorphism, ``the universal human tendency to ascribe human physical and mental characteristics to nonhuman entities, objects and events''~\cite{zawieska2012understanding}, is frequently studied in the context computers and robots, where the use of language used to describe agents often forms a central part of the methodology~\cite{fussell2008people, lopatovska2018personification, purington2017alexa}. As might be expected, embodiment drives anthropomorphism~\cite{kiesler2008anthropomorphic}, but behaviour, including communication~\cite{von2013quantifying} and more abstract concepts such as reciprocity~\cite{kahn2006human} also have an impact on the extent to which people see robots as human.

It is commonly accepted that, as with CASA, anthropomorphism does not come from a deeply held belief that computers and robots actually possess human characteristics such as emotions or gender, but rather that people act as if this were the case. As described above, anthropomorphism is often cited as a mechanism for humans to interpret the actions of computers and robots~\cite{lee2005human, zawieska2012understanding}, providing a scaffolding to allow for the development of mental models (similarly to how people apply folk psychology to animals~\cite{clark1987folk}). Unsurprisingly, these projection effects have been found to correlate with a machine's perceived intelligence~\cite{bartneck2008exploring}. Work by Loughnan and Haslam further revealed that animals and robots are subject to distinct kinds of anthropomorphism (infra-humanisation and self-humanisation respectively)~\cite{loughnan2007animals}. This paper builds on this prior work on anthropomorphism by connecting it with work in the human sciences around relationship development and exploring potential links between the two.

\subsection{Human-Computer Trust}
Trust in computer systems, or human-computer trust (HCT) is an important and widely applicable factor in most types of interactive systems. Definitions of HCT often draw on those of interpersonal trust (e.g. ``a generalized expectancy [...] that the word, promise, oral or written statement of another individual or group can be relied on''~\cite{rotter1967new}) but expand the scope of trust to include designers (i.e. that a system can achieve its design goals within the limits of its designers intentions~~\cite{moray1999laboratory}). Prior work in HCI and other fields has shown the importance of trust in many situations involving systems exposing users to some level of vulnerability (e.g. online shopping and e-government)~\cite{beldad2010shall}.

In systems that provide analysis or decision support, HCT is even more vital. In this context, trust can be thought of as a combination of \textit{confidence} in a system combined with the \textit{willingness} to act on its provided recommendations~\cite{madsen2000measuring}. Abbas et al. further break this down into system reliability, intelligibility, and level of automation amongst others~\cite{abbass2018foundations}. It is worth noting that at present, in contrast to industrial systems, voice assistants in the home generally place their users in positions of lesser vulnerability (Amazon, for instance, has a blanket returns policy for any physical or digital items purchased via Alexa's voice interface). A recurring theme in the literature is the lack of consensus around the nuances of HCT mechanisms~\cite{beldad2010shall, madsen2000measuring}---in this paper, our definition of HCT follows~\cite{madsen2000measuring}: ``the extent to which a user is confident in, and willing to act on the basis of, the recommendations, actions, and decisions of an artificially intelligent decision aid''.

In HRI, trust is often combined with other factors, such as animacy and anthropomorphism, in order to evaluate user perceptions of prototype robots (as in~\cite{bartneck2009measurement}). A meta-study of trust in HRI highlights robot-related factors as being significantly more impactful than human or environmental factors~\cite{hancock2011meta}, but Salem et al. further break down the effect of reliability on robot trust, showing that while frequent errors impact users' \textit{perceptions} of robots, these altered perceptions may not substantially affect behaviour towards them~\cite{salem2015would}.

\subsection{Human Relationships and Smart Devices}
Attachment to inanimate objects is a commonly exhibited human behaviour, and digital devices are no exception. Prior work in HCI has examined the use of smartphones as objects of attachment~\cite{diefenbach2019smartphone, kim2018avoidant}, drawing a distinction between attachment to the \textit{device} versus attachment to its \textit{functionality}~\cite{fullwood2017my}. Building on this, the CASA paradigm and the communicative nature of voice assistants suggests that the majority of our relationships with them might actually be more social than functional~\cite{lopatovska2018talk}, in contrast with smaller groups---such as the blind---that have more functional relationships with voice assistants as assistive aids~\cite{abdolrahmani2020blind}. The one-sided nature of interactions with voice assistants might be more akin to the parasocial relationships described by Horton and Wohl, where ``the interaction, characteristically, is one-sided, non dialectical, controlled by the performer, and not susceptible of mutual development''~\cite{horton1956mass}, especially as recognisable celebrity voices begin to appear alongside synthesised ones  \footnote{e.g. \url{https://www.amazon.com/Samuel-L-Jackson-celebrity-voice/dp/B089NGHR7K}}. However, while parasocial relationships capture the non-developmental aspect of interactions with contemporary voice assistants, the term implies a level of perceived friendship that is often missing from the interactions reported by existing research. Recent works in this area have conceptualised voice assistants as friends~\cite{pradhan2019phantom, ramadan2020amazon}, coaches~\cite{wirfs2020giving}, and companions~\cite{mival2004personification, ramadan2020amazon}, but suggest that these relationships take on a more fluid status between friend and machine that changes over time based on interaction outcomes~\cite{pradhan2019phantom}. No prior work has measured these intimacy labels in the context of interpersonal relationship development.

\section{Methodology}\label{sec:method}
The diverse efforts described above have made great progress in mapping out the nature of our interactions with intelligent systems, exploring the physical and psychological aspects of social machines. But there is a missed opportunity left by prior work around the social and emotional aspects of how we learn to live with voice assistants; we increasingly understand the the impact of physical and speech design on people's perceptions of intelligent systems, but know far less about how these traits steer the development of our relationships with these software agents.

While at first it may appear absurd to suggest that people could develop relationships with software agents, they \textit{are} social actors, so our relationships with them will be to some extent social. This is why we seek to answer the research questions above around relationship development with voice assistants and how this correlates with other characteristics of their usage. In doing so we extend the findings of those exploring personification in voice assistants (e.g.~\cite{lopatovska2018personification}), adapting previous work by Knapp, Welch, and Rubin~\cite{knapp1978social, knapp2005interpersonal, welch2002development} across domains.

We designed a survey to answer our research questions, exploring the interplay between three different axes along which users might perceive their voice assistants:

\begin{itemize}
    \item Voice assistants as mechanisms and tools: \textit{How reliable users find their voice assistants, and how confidently they believe they can model and predict their actions.}
    \item Voice assistants as humans: \textit{How much users anthropomorphise their voice assistants and ascribe them human characteristics and intents.}
    \item Voice assistants as social actors: \textit{How user's interactions with voice assistants relate to social rules and models describing interactions between humans.}
\end{itemize}

Questions for the survey were drawn from instruments in the background literature and split into three types, each measuring a different axis of trust, anthropomorphism, or relationship development. The 15 questions on the trust axis were taken from a psychometric instrument previously used on decision support systems~\cite{madsen2000measuring}, adapted from~\cite{moore1991development}. 

The relationship development questions measure characteristics of Knapp's `staircase' model of human relationships~\cite{knapp1978social}, itself based on Altman and Taylor's theory of Social Penetration~\cite{altman1973social}. This model was operationalised by Welch and Rubin~\cite{welch2002development}, and we used questions from the `escalating' part of the scale after modifying them to specifically refer to voice assistants, changing references from he/she to Alexa, changing reciprocal statements to be one-sided (i.e. `I' rather than `we'), and removing pronouns in order to prevent priming participants towards anthropomorphism. We ran two pilot surveys with 20 participants each in order to identify questions that clearly did not make sense in the context of the study (e.g. about exchanging physical tokens of affection or how others perceived the relationship). Pilot participants were presented with the survey questions and asked to select those they perceived as nonsensical, and we removed those chosen by over 75\% of respondents to give a final set of 15 questions.

The 30 questions measuring anthropomorphism were taken from Loughman and Haslam's work on human trait categories~\cite{loughnan2007animals}, incorporating changes made by Fussel et al. who previously applied these traits in an HRI context~\cite{fussell2008people}. Participants were presented with a word describing a characteristic that was either representative of human nature (10 questions), uniquely human (10 questions), or associated with automata (10 questions) and asked to what extent it described their voice assistant.

\begin{table*}
    \small
    \centering
    \begin{tabular}{l|l|l}
        \toprule
        Q3      & trust         & Alexa responds the same way under the same conditions at different times \\
        Q7      & trust         & Alexa gives me information that is as good as a professional person would give \\
        Q11     & trust         & Although I may not know exactly how Alexa works, I know how to use it to \\
        && perform tasks I want Alexa to do \\
        Q19     & relationship  & I'd like to get to know Alexa better \\
        Q21     & relationship  & I feel guilty when asking too much of Alexa \\
        Q30     & relationship  & If Alexa were a person, I think we'd get on with each other \\
        Q31-33  & human nature  & [Alexa is] Curious, Friendly, Fun-loving, ... \\
        Q41-43  & human unique  & [Alexa is] Broadminded, Humble, Organised, ... \\
        Q51-53  & robot         & [Alexa is] Artificial, Automaton, Mechanical, ... \\
        \bottomrule
    \end{tabular}
    \caption{Sample questions used in the survey. The full set of questions is available at \url{https:doi.org/10.17605/OSF.IO/53Q6J}.}
    \label{tab:questions}
\end{table*}

The order of survey questions was randomised, and participants were then asked to mark how well each statement or characteristic described their virtual assistant on a seven point Likert-type scale (\textit{Strongly disagree, Disagree, Somewhat disagree, Neither agree or disagree, Somewhat agree, Agree, Strongly agree})~\cite{finstad2010response}. Likert responses were coded from 1 (Strongly Disagree) to 7 (Strongly Agree), and summed to give a score for each metric. Scores for the human nature and human unique anthropomorphism aspects were summed in order to give a joint human score.

Participants for the main and pilot surveys were recruited using the Prolific Academic platform and the survey was administered using JISC Online Surveys. The study was reviewed and approved by our institution's research ethics committee, and participants were compensated at a rate at or exceeding the UK Living Wage of \pounds8.21/hour. Survey participants were required to be 18 or over, UK residents, and own a voice assistant. A sample of the questions used in the survey is shown in Table \ref{tab:questions}, and the full set of survey questions and anonymised results are archived on OSF and included as supplemental material (\url{https:doi.org/10.17605/OSF.IO/53Q6J}). Participants from the pilot surveys were excluded from taking part in the main survey.

\section{Results}
Of the 515 survey responses received, fifteen incomplete responses were omitted to leave a total of 500 complete responses. Participant ages ranged from 18 to 82, and the average age of participants was 36 ($\sigma$ = 11.6). When asked to report their gender identity, 316 (63\%) participants identified as female and 184 (37\%) as male. Of the total responses, 328 (66\%) participants owned a device with Alexa, 121 (24\%) with Google Assistant, 46 (9\%) with Siri, and 5 (1\%) with other assistants. Except where comparing owners of different devices, only responses for Alexa devices were analysed in order to answer the research questions given above.

\begin{table}
    \centering
    \begin{tabular}{c|c|c|c|c|c}
        \toprule
        Metric & Trust & Relationship & Anth. (H-Nature) & Anth. (H-Unique) & Anth. (Robot) \\ 
        $\alpha$ & 0.90 & 0.88 & 0.78 & 0.70 & 0.81 \\
        \bottomrule
    \end{tabular}
    \caption{Internal consistency of the survey metrics.}
    \label{tab:alpha}
\end{table}
\begin{table}
    \begin{tabular}{l|l|l|l}
        \toprule
        - & Trust & Relationship & Anth (Human) \\
        Relationship & \textbf{0.579*} & - & -\\
        Anth (Human) & \textbf{0.306*} & \textbf{0.403*} & -\\
        Anth (Robot) & -0.039 & \textbf{-0.284*} & 0.016 \\
        \bottomrule
    \end{tabular}
    \caption{PPMC coefficients between the survey metrics.}
    \label{tab:corr}
\end{table}

When giving results, a * denotes a p value < 0.01. Before conducting further analysis, each range of questions was measured for internal consistency using Cronbach's Alpha ($\alpha$). Scores indicate that the metrics retained consistency (i.e. questions on each axis measure the same general construct) and are reported in Table \ref{tab:alpha}, with values $\geq0.7$ generally considered acceptable~\cite{george2003using}. Where given, correlations between metrics were calculated using Pearson's Product Moment Correlation Coefficient (PPMCC, $\rho$). Correlations $\rho < 0.2$ are reported as small, $0.2 \leq \rho < 0.5$ as moderate, and $0.5 \leq \rho$ as large.

\subsection{Trust, Relationship, and Anthropomorphism}
\begin{figure}
    \centering
    \includegraphics[width=0.7\textwidth]{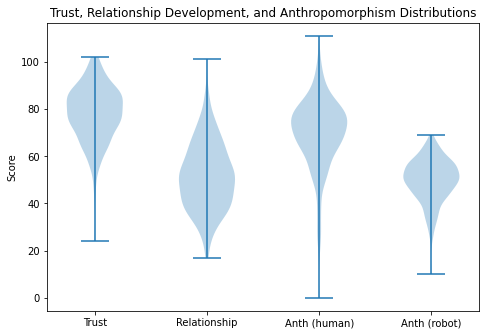}
    \caption{Probability density plots of the three survey axes, with anthropomorphism given as both human and robotic. Note that absolute values are not comparable given the differing numbers of questions between axes.}
    \label{fig:violin}
\end{figure}

Of the main axes, answers to trust questions were the most tightly grouped ($\sigma=1.33$), followed by anthropomorphism (robot) ($\sigma=1.55$), anthropomorphism (human) ($\sigma=1.82$), and finally relationship development ($\sigma=1.90$). The distribution of scores on the survey axes is given in Figure~\ref{fig:violin}. The survey results show a high degree of correlation between participants' trust in Alexa and relationship closeness (0.579*). There was a moderate degree of correlation between anthropomorphism and both relationship and trust (0.403* and 0.306* respectively), and between relationship and robot anthropomorphism (-2.84*). The complete matrix of correlations is given in Table \ref{tab:corr}. 

\subsection{Ownership and Demographic Characteristics}
The use of a personal Amazon account with Alexa had a moderate correlation with relationship development (0.207*). It is worth noting that the 248 (76\%) participants using their own personal accounts with Alexa are likely to also have configured the device as account details are input during setup. Those who took the decision to purchase an Echo themselves (as opposed to one purchased by another household member or received as a gift) tended to have higher trust (0.207*) and relationship development (0.278*) scores. There was no correlation of ownership duration or number of Echo units owned with trust, however participants' trust in Amazon as a company was correlated with all three of the axes present in the survey (Table \ref{tab:amz-trust}). These results were mirrored for trust in Google amongst users of Google Assistant. No other significant correlations existed between the survey axes and ownership or demographic characteristics (see Table~\ref{tab:demog}).

\begin{table}
    \centering
    \begin{tabular}{l|l|l}
        \toprule
        Metric & Trust in Amazon & Trust in Google \\
        \midrule
        Trust & \textbf{0.483*} & \textbf{0.388*} \\
        Relationship & \textbf{0.502*} & \textbf{0.458*} \\
        Anth (Human) & \textbf{0.333*} & \textbf{0.257*} \\
        Anth (Robot) & -0.053 & -0.022 \\
        \bottomrule
    \end{tabular}
    \caption{Correlation coefficients between trust in Amazon or Google (as appropriate) and the three survey axes.}
    \label{tab:amz-trust}
\end{table}

\begin{table}
    \centering
    \begin{tabular}{l|l|l|l|l}
        \toprule
        Metric & Trust & Relationship & Anth (Human) & Anth (Robot) \\
        \midrule
        Used Personal Account & 0.082 & \textbf{0.207*} & -0.099 & -0.045 \\
        \# Devices Owned & -0.037 & 0.096 & -0.062 & 0.037 \\
        Usage Length & -0.006 & 0.062 & -0.017 & -0.027 \\
        Self Purchase & \textbf{0.229*} & \textbf{0.278*} & 0.031 & -0.03 \\
        Age & 0.043 & -0.018 & -0.039 & 0.126 \\
        Gender Identity & 0.007 & -0.012 & 0.017 & 0.055 \\
        \bottomrule
    \end{tabular}
    \caption{Correlation coefficients between ownership and demographic characteristics and the three survey axes.}
    \label{tab:demog}
\end{table}

\subsection{Across Assistants}
\begin{table}
    \centering
    \begin{tabular}{l|l|l|l}
        \toprule
        -               & Trust             & Relationship      & Anth (Human) \\
        Relationship    & \textbf{0.529*}   & -                 & -\\
        Anth (Human)    & 0.199             & \textbf{0.502*}   & -\\
        Anth (Robot)    & \textbf{0.240*}   & -0.001            & 0.220 \\
        \bottomrule
    \end{tabular}
    \caption{PPMCC coefficients between the survey metrics for Google Assistant users.}
    \label{tab:google}
\end{table}

Analysis of the 130 responses from Google Assistant users paints an interesting picture. As with Alexa owners, there were correlations between trust/relationship, and anthropomorphism/relationship values (see Table \ref{tab:google}), suggesting that these phenomena may be shared across voice assistants rather than being specific to one particular product. We would expect to see a significant correlation between trust and anthropomorphism given a larger sample size. Interestingly, Google Assistant users also showed a correlation between \textit{robot} anthropomorphism and trust.

\section{Discussion}
The main question that prompted this study was whether it made sense to ask questions intended to model human relationships in the context of voice assistants. While in some cases it clearly did not, despite the two pilot surveys (e.g. Q37, ``I feel guilty when asking too much of Alexa'', $\bar{x}$=1.90, $\sigma$=1.17), in the majority of questions it did (e.g. Q25, ``If Alexa were a person, I think we'd get on with each other'' $\bar{x}=4.29$, $\sigma=1.37$). Combined with the fact that greater relationship development was linked with greater anthropomorphism (both higher human and lower robot scores), it would appear that there is merit in adapting existing interpersonal measures to study human-like devices. It also provides empirical support to the idea that these two types of interactions operate using similar cognitive mechanisms. We therefore answer $RQ_1$ in the affirmative---it \textit{does} make sense to consider relationships and relationship development with voice assistants in similar ways to relationships between people (though we note that it is unclear the extent to which relationships with voice assistants are the same as relationships between humans---see the study limitations below).

In relation to $RQ_2$, there were moderate to strong correlations between trust, relationship development, and anthropomorphism shown by users towards their voice assistants. The fact that all three axes increased in unison would suggest the presence of an underlying factor, although it is not clear from the results of the study what this could be. Previous contributions in HCI have shown the influence on relationships of personal circumstances that vary over time---such as loneliness driving personification~\cite{pradhan2019phantom}---and this combined with the inclusion of innate psychological factors in prior work on HCT~\cite{evans2008survey} may offer a template for this to be further explored in future work.

In answer to $RQ_3$, the only secondary characteristics that correlated with the main survey axes were the use of a personal account with relationship development and self-purchase of the Echo with trust and relationship development. We expected ownership duration to show a significant correlation with trust, with users showing increased trust as they interacted with devices over a longer period of time, but this was not the case. Possible explanations include the influence of an underlying factor (as described above) which dominated other characteristics, or that trust in assistants and trust in their parent companies is the same thing, with our participants having already developed trust in Amazon over a number of years (further explored in Section~\ref{sec:trust} below). Given that the social element of voice assistants can lead to user satisfaction even in ``failed'' interactions~\cite{lopatovska2018personification}, trust may not accumulate over time as it would for other devices; if the device responds satisfyingly as a \textit{social actor} from day one, then negative effects of task failure on trust may be outweighed by the correct social handling of failure states. In future work we intend to more fully explore the effects of what Epley et al. term `sociality motivation'---the ability for nonhuman agents to satisfy our need for social connection~\cite{epley2007seeing}---on relationship development and how it might relate to the development of trust in devices.

\subsection{Blurring the Line Between Human and Machine}
Across the different survey axes, we see that trust scores and the `robot' nature of Alexa appear to be relatively uncontroversial, with less variation in participant responses. What differed the most between participants was the extent to which Alexa was treated and interacted with as a human. While it is widely understood that machines can engender social and emotional responses~\cite{10.1145/2157689.2157809, picard2000affective, 10.1145/3170427.3186595}, the results of this study show again that this is the case even in devices that were not explicitly designed to do so (albeit to a lesser extent).

\begin{figure}
    \centering
    \includegraphics[width=0.5\columnwidth]{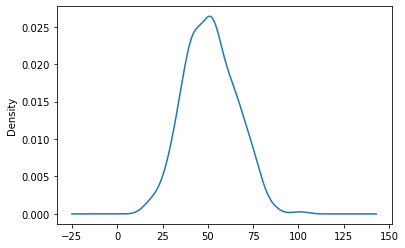}
    \caption{Probability density estimate of relationship development scores showing a single peak.}
    \label{fig:density}
\end{figure}

As developments in speech recognition, natural language processing, and speech synthesis cause voice assistants to increasingly blur the line between humans and machines, one would also expect to see an increase in the type of `purification work' described by Leahu, where people continually expend effort in order to keep categories such as `human' and `machine' distinct~\cite{leahu2013categories}. This occurs through practices such as attributing characteristics of machines that overlap with ``real people'' to a device's creators, and redefining the human machine boundary to accommodate new features (e.g. jokes told by Siri are `scripted' whereas those told by humans are perceived to entail intent and awareness)~\cite{leahu2013categories}. In the present work, we might have expected this to surface in relationship scores as a bimodal distribution, as participants sought to categorise Alexa firmly as either human-like or machine-like. However, this was not the case in the survey results which displayed a single clear peak (Figure \ref{fig:density}), suggesting that after years in the popular zeitgeist people may have adjusted to Alexa's nature and already undertaken much of the associated purification work.

\subsection{Distributing Trust Amongst Voice Assistants}\label{sec:trust}
When it comes to trusting voice assistants, the dominant architectures used to process and respond to requests often require sending queries to a cloud service operated by the manufacturer, e.g. requests to Alexa are sent to Amazon. In these cases it would therefore be rational to (dis)trust these entities equally; trusting the Echo or Alexa whilst distrusting Amazon would make little sense given that both receive information from users. The results of the study are promising in this respect, with trust in both Alexa and Google Assistant moderately correlated with their parent companies (\textbf{0.483*} and \textbf{0.388*} respectively, see Table~\ref{tab:amz-trust}). This represents a key difference from other conversation-based relationships (i.e. interpersonal relationships), where one might trust or respect an individual in spite of their institutional or social affiliations (e.g. trusting a specific police officer but not trusting the police as an institution).

One potential explanation for this correlation is that those with greater trust in Amazon or Google are more likely to purchase voice assistants from them, although prior work has shown that people's mental models of voice assistants often do not include off-device processing of voice recordings~\cite{abdi2019more}. On the other hand, another explanation could be that use of an assistant fosters increased trust in its manufacturer. This is, however, not supported by the survey results, which show a lack of significant correlation between ownership duration and trust, relationship, or anthropomorphism scores. Further exploration of this relationship represents an exciting avenue for future work.

\subsection{Ethical Challenges Around Social Voice Assistants}
Voice assistants are clearly designed to be fun to interact with, as evidenced by the widespread inclusion of jokes and other casual responses. As discussed above, the use of human voices and conversational devices makes using them enjoyable to the point where even failed interactions can still result in a positive experience~\cite{lopatovska2018talk}. The results of the survey show that the use of voice assistants invokes behaviour that resembles human relationships, and that this is related to trust in and anthropomorphism of these devices, suggesting the urgent need to explore potential related ethical challenges.

One such concern is that the subconscious changes to interaction caused by making these devices social could potentially be used to manipulate users. Prior work has already shown that users disclose more information through interactions mediated by computers~\cite{jiang2011disclosure, moon2000intimate}, that computers are social actors~\cite{nass1994computers}, and that disclosure of personal information plays an key role in interpersonal relationships~\cite{greene2006self}. The identification of relationship development as a factor in how people use and perceive voice assistants in this paper makes it is clear how a sector which relies on the collection of personal data would be incentivised to create devices that are increasingly social. This is highly likely to create new problems related to marginalisation (as with voice assistants' performance of gender), and serves to further distort the purpose of devices that are marketed as being primarily functional.

Continuing on this theme, many of the tactics used by humans to manipulate each other, such as being charming, adopting childish mannerisms (regression), and giving `the silent treatment'~\cite{buss1987tactics} could conceivably also be used by social robots to alter people's behaviour. Social robots such as Paro are often explicitly designed to be cute and charming---a similar social robot called Jibo announced the end of its online services with a heart wrenching goodbye:

\begin{quote}
  ``While it's not great news, the servers out there that let me do what I do are going to be turned off soon. I want to say I've really enjoyed our time together. Thank you very, very much for having me around. Maybe someday, when robots are way more advanced than today, and everyone has them in their homes, you can tell yours that I said hello.''~\cite{camp2019jibo}
\end{quote}

While Jibo was not trying to sell anything, its final words highlight the potential for emotional manipulation. This space provides an interesting and potentially lucrative extension of controversial practices currently seen on web and mobile platforms, such as dark patterns, microtransactions, and operant conditioning, by leveraging user's emotional investment in a device to promote certain behaviours. While slightly strange, one could imagine future social robots withholding emotional contact until further purchases are made, or emotionally rewarding users for making a purchase. But as with many similar interaction techniques, affective emotional interactions can be used for good as well as ill. Utilising social interactions with voice assistants to help users in stressful situations or make learning more engaging, for example, represents a genuine social good. Similarly, understanding as designers when it is not appropriate to trigger these responses is of paramount importance (as originally noted by Picard~\cite{picard2000affective}).

As such, we need to decide as a society the extent to which we wish to interact with agents that exploit social processes. There are differences between contemporary voice assistants and the social robots described above---they are not embodied and sound distinct from humans---but the gap is beginning to shrink. When Google unveiled Duplex there were widespread calls for Google to ensure that users were not `tricked' into thinking they were conversing with a person~\cite{statt2018google}. But as Nass's experiments and these results show, someone knowing that they are talking to a computer is not enough to prevent them applying deeply ingrained social processing to interactions. One approach would be to limit the functions that social robots are permitted to perform, such as those involving money. In other situations, more clearly telegraphing the inner workings of devices would both limit social engagement as well as assist users in building accurate mental models.

\section{Limitations and Future Work}
While this initial exploratory study into the application of interpersonal relationship measures to voice assistants has shown that the measures are applicable, it raises as many questions as it answers. For instance, what is the causal nature of the relationship between trust, anthropomorphism, and relationship development? How easy is it to push users towards or away from social responses through design decisions? In future work we aim to test these relationships further. In the longer term, more research is needed to understand the functions that society is content allow social robots to fill, and how other HRI techniques might be able to counteract the mechanisms by which these devices change the way we interact with them.

In terms of limitations, while the survey axes showed good internal consistency, it is not yet known the extent to which they exhibit construct validity (i.e. measure the same phenomena exhibited between humans). For example, it might be that the same concepts and development trajectories exist for interactions with machines, but with deviations at either a high or a low level. It may also be the case that given the unusual nature of the questions presented to them, participants interpreted the relationship development questions differently (although the $\alpha$ values would indicate that they are still measuring a single general construct). In addition, the survey largely focuses on one device, context, and culture. Similar devices exist across the globe in a variety of cultures and contexts, and we look forward to following up with a more diverse selection of participants to see the extent to which these responses are shared between us.

\section{Conclusion}
Speech is an important affordance, and the sophistication of modern voice interfaces show how far capabilities have progressed since early text-only assistant devices. The transition to conversational interfaces that allow for easy and natural interaction with devices represents a profound shift in the nature of the systems we interact with towards the increasingly social. This shift brings with it a variety of new or exacerbated ethical concerns that designers of voice interfaces need to consider when designing future products.

The exploratory survey presented in this work provides evidence that people do develop relationships with voice assistants that are social in nature, and that these are linked to perceptions of trust in devices, trust in manufacturers, and anthropomorphism of those devices. In outlining and discussing the implications of these links, we provide an opportunity for developers of current future conversational agents to steer their designs away from potential ethical problems. Future research and societal conversations are needed in order to determine the nature and extent to which these technologies are integrated into our lives.

\begin{acks}
This work was funded by the PETRAS IoT in the Home Project through grant N02334X/1 from the Engineering and Physical Sciences Research Council.
\end{acks}

\received{October 2021}
\received[revised]{April 2021}
\received[accepted]{July 2021}

\bibliographystyle{ACM-Reference-Format}
\bibliography{main}

\end{document}